\documentclass[12pt]{article}
\usepackage[dvips]{graphicx}
\usepackage{amsmath}

\renewcommand{\bar}[1]{\overline{#1}}
\newcommand{\half}{{\frac{1}{2}}}
\newcommand{\threehalf}{{\frac{3}{2}}}
\newcommand{\fivehalf}{{\frac{5}{2}}}
\newcommand{\sevenhalf}{{\frac{7}{2}}}
\newcommand{\ninehalf}{{\frac{9}{2}}}
\newcommand{\elevenhalf}{{\frac{11}{2}}}
\newcommand{\thirteenhalf}{{\frac{13}{2}}}

\def\Dslash{\raise.15ex\hbox{/}\kern-.7em D}
\def\Pslash{\raise.15ex\hbox{/}\kern-.7em P}

\begin {document}
\begin{flushright}
{\small
SLAC--PUB--10693\\
September 2004}
\end{flushright}

\medskip

{
\begin{center}
{{\bf\LARGE Baryonic States in QCD From\\ \vspace{0.3cm} 
      Gauge/String Duality at Large ${\bf N_C}$}\footnote{Work supported by the Department
of Energy contract DE--AC02--76SF00515.}}\\
\vspace{20pt}
{\bf Guy F. de T\'eramond\footnote{E-mail: gdt@asterix.crnet.cr}
and Stanley J. Brodsky\footnote{E-mail: sjbth@slac.stanford.edu} } \\
\vspace{10pt}
{\it Universidad de Costa Rica, San Jos\'e, Costa Rica\\ \vspace{10pt}
Stanford Linear Accelerator Center, \\ 
 Stanford University, Stanford, California 94309, USA 
\vspace{10pt}
 }
\end{center}
}

\medskip

\begin{abstract}

We have computed the  baryon spectrum in the context of $\mathcal{N} = 4$
super-conformal Yang-Mills theory using AdS/CFT duality.
Baryons are included in the theory by adding an open string sector,
corresponding to quarks in the fundamental and higher
representations.
The hadron mass scale is introduced by imposing boundary conditions 
at the wall at the end of AdS space. 
The quantum numbers of each baryon, 
are identified by matching the fall-off of the string wavefunction 
$\Psi(x,r)$ at the asymptotic $3+1$ boundary to the
operator dimension of the lowest three-quark Fock state, subject
to appropriate boundary conditions. 
Higher Fock states are matched quanta to quanta with quantum 
fluctuations of the bulk geometry about the fixed AdS background,
maintaining conformal invariance.
The resulting four-dimensional 
spectrum displays a remarkable resemblance to the physical baryon
spectrum of QCD, including the suppression of spin-orbit interactions. 

\end{abstract}

\smallskip

\begin{center}
{\it Presented at  \\
ETC Workshop on Large $N_C$ QCD 2004\\
5-9 July 2004  \\
Trento, Italy \\
 }
\end{center}

\vfill

\newpage

\parindent=1.5pc
\baselineskip=16pt

\setcounter{footnote}{0}

\section{Introduction}

~~~~~An outstanding  consequence of Maldacena's 
duality~\cite{Maldacena:1997re} between 
10-dimensional string theory on $AdS_5 \times S^5$
and Yang-Mills theories at its conformal 3+1 
spacetime boundary~\cite{Gubser:1998bc,Witten:1998qj}
is the potential to describe processes for physical
QCD which are valid at strong coupling and 
do not rely on perturbation theory. As shown by  Polchinski
and Strassler~\cite{Polchinski:2001tt}, dimensional counting
rules~\cite{Brodsky:1973kr} for the leading 
power-law fall-off of hard exclusive glueball scattering
can be derived from a
gauge theory with a mass gap dual to supergravity in warped spacetimes. The modified theory
generates the hard behavior expected from QCD, instead of the soft
behavior characteristic of strings. Other examples are the description
of form factors at large transverse momentum~\cite{Polchinski:2001ju} and
deep inelastic scattering~\cite{Polchinski:2002jw}. The discussion of
scaling laws in warped backgrounds 
has also been addressed in~\cite{Boschi-Filho:2002zs,Brower:2002er,Andreev:2002aw}.

The AdS/CFT duality gives a non-perturbative definition
of quantum gravity in a curved background which is asymptotic to a product 
of Anti-de Sitter space AdS and a compact Einstein space  $X$.
As originally formulated~\cite{Maldacena:1997re}, a correspondence was
established between the supergravity approximation to Type IIB string theory,
and the large $N_C$ brane decoupling limit with gauge dynamics
corresponding to $\mathcal{N} = 4$
super Yang-Mills (SYM) in four dimensions. The bulk geometry has exact conformal
geometry $AdS_5 \times S^5$ in the near-horizon region $r \ll R$, where
$R = ({4 \pi g_s N_C})^{1/4} \alpha'^{1/2}$ is  the radius of AdS and the
radius of the five-sphere. 
The extra five dimensions of $S^5$
correspond to the $SU(4)$ global symmetry which rotates the particles
present in the SYM supermultiplet in the adjoint representation of $SU(N_C)$. 
The conformal group $SO(2,4)$ is identified with the isometry group of
$AdS_5$, and $SU(4) \simeq SO(6)$ with the isometries of $S^5$.  The supergravity duality requires
a large value of $R$ corresponding to a large value of the 't Hooft
parameter, $g_s N_C$~\cite{Aharony:1999ti}.

In a recent attempt to extend the glueball results to 
hadrons~\cite{Brodsky:2003px}, we used  the
AdS/CFT correspondence to determine the basic properties of hadronic
light-front wavefunctions (LFWF) in QCD based on the underlying
conformal symmetry of the duality for 
pointlike hard-scattering processes
which occur in the large-$r$ region of AdS space. 
The scaling behavior of the string modes determine the behavior
of the  QCD hadronic wavefunction, giving a precise counting rule
for each Fock component state with an arbitrary number of quarks and gluons 
and internal orbital angular momentum\footnote
  {In~\cite{Brodsky:2003px} we examined the
  possibility of identifying the
  internal orbital momentum of hadrons with Kaluza-Klein excitations
  of the internal space $S^5$. Henceforth we follow the interpretation
  given here in terms of quantum fluctuations about the AdS background.}.
The discussion
was carried out in terms of the lowest dimensions of interpolating fields
near the boundary of AdS, treating the boundary values of the string states $\Psi(x,r)$
as a product of quantized operators which create $n$-partonic states out of the
vacuum~\cite{Brodsky:2003px}. Our AdS/CFT derivation validate QCD perturbative results and confirm the
dominance of the quark interchange mechanism~\cite{Gunion:1972gy} for
exclusive QCD processes at large $N_C$. The
predicted orbital dependence coincides with the fall-off of 
light-front Fock wavefunctions derived in perturbative
QCD~\cite{Ji:2003fw}. Since all of the Fock states of the LFWF beyond
the valence state are a manifestation of quantum fluctuations,
it is natural to match quanta to quanta the
additional dimensions with the metric fluctuations of the bulk
geometry about the fixed AdS background.

For large values of $R$, or small curvature of AdS space, it is expected that the dual 
of a Yang Mills theory is classical gravity. 
The correspondence also implies that the
dual of strongly coupled QCD is a weakly coupled string.
Since QCD is weakly coupled at high energies, the dual theory is expected to be a strongly 
coupled string model at small 't Hooft coupling and
would require the understanding of strings in highly curved
backgrounds, extending the semiclassical approximation to include
quantum effects on the string theory side.
The behavior of string states in the infrared  region
is dependent on dynamics at small-$r$, and it is a priori unknown.
Non-conformal aspects are needed to make contact with the
real world. The introduction of quarks in the
fundamental representation is also crucial, requiring an
open string sector. 

 In spite of the difficulties mentioned above, important progress
has been achieved by extending the AdS/CFT correspondence beyond the
supergravity approximation
to construct string duals to non-conformal gauge 
theories\footnote{For a review see~\cite{Bigazzi:2003ui}. See also~\cite{Aharony:1999ti}.}.
Even if the detailed form of the metric at small-$r$ is unknown, salient QCD dynamical 
features, such as the generation of a mass gap and a hadron spectrum, will follow from 
the deformation of the AdS conformal background at small $r$. Indeed
from~\cite{Polchinski:2001tt} there follows a simple relation between
the 10-dimensional string
scale $\alpha'_s$  and the Yang-Mills 4-dimensional scale $\alpha'_{QCD}$ in a warped space:
$\alpha'_{QCD} \sim \alpha'_s \left( R / r_o \right)^2$,
where the cutoff $r_o = \Lambda_{QCD} R^2$, breaks conformal invariance
and allows the introduction of the QCD scale. 

A physical hadron in four-dimensional Minkowski space has four-momentum $P_\mu$ and hadronic
invariant mass states given by $P_\mu P^\mu = \mathcal{M}^2$. 
The string wavefunction in $r$ is the extension
of the baryon wavefunction into the fifth dimension: we thus
analytically match the three-quark proton wavefunction in $3 + 1$
space to its corresponding string wavefunction using the
10-dimensional wave equation. Different values of $r$ correspond to
different energy scales at which the hadron is examined. In particular,
the $r \to \infty$ boundary corresponds to the $Q \to \infty$, zero separation limit.  
The physical string modes
\begin{equation}
\hat\Psi \equiv \langle 0 \vert \hat\Psi \vert P \rangle 
\sim  e^{- i P\cdot x}~ f(r)~ Y(y),
\end{equation}
are plane waves along the Poincar\'e
coordinates, and $Y$ is a function of the transverse coordinates $y$.
For large-$r$, $f(r)$ scale as $f(r) \sim r^{-\Delta}$, 
where $\Delta$ is the conformal dimension of the string state, the
same dimension of the interpolating operator which creates a hadron
out of the vacuum~\cite{Polchinski:2001tt}.
The string modes are coupled to the matter fields 
of the conformal theory as determined by the boundary limit 
between the string partition function and
the generating functional of the quantum field
theory~\cite{Gubser:1998bc,Witten:1998qj}. For example,
the quantum numbers of each baryon, including intrinsic spin 
and orbital angular momentum, are
determined by matching the dimensions of the string modes $\Psi(x,r)$,
with the lowest dimension of the baryonic interpolating operators 
in the conformal limit. 
Although the underlying string theory dual to $QCD(3+1)$ is unknown, 
the introduction
of quark fields in the fundamental representation and the
symmetries in the asymptotic boundary, should provide the
conditions required to establish a precise matching 
between the string modes in the semiclassical gravity
approximation and boundary states with well defined number of partons. In particular,
the known light baryons are dual to spin-$\half$ and spin-$\threehalf$
strings and consequently there is not a vastly large mass gap between 
baryons with total angular momentum $J \le 2$ and  $J > 2$
for large $N_C$. 

After stating in Sec. 2 some basic properties of the correspondence between
string modes in AdS space and baryon states at the asymptotic
boundary, we  confront our results with the spectrum of 
nucleon and $\Delta$ orbital 
resonances in Sec. 3. Some concluding remarks are given in Sec. 4.

\section{Baryon Interpolating Operators and String Modes}

~~~~~A precise statement of the duality between a string/gravity theory on a $(d+1)$-dimensional
Anti-de Sitter space $AdS_{d+1}$ and the large $N_C$ limit of a conformal theory at its
$d$-dimensional boundary, is given formally in terms of the full partition function of the string
theory in the bulk $Z_{string}$ which should coincide with the generating functional of the
conformal field  theory $Z_{CFT}$ on the AdS
boundary~\cite{Gubser:1998bc,Witten:1998qj}:
$Z_{string}\left[\Psi(x,z=0) \right] =
Z_{CFT}\left[\Psi_o\right]$.
For spin-$\half$ dilatino in the bulk, the duality involves positive and
negative chirality components~\cite{Muck:1998iz} $\Psi^\pm =
\half (1 \pm \gamma_5) \Psi$,
which couple with CFT operators
$\mathcal{O}^+$ and $\mathcal{O}^-$
\begin{equation}
Z_{CFT}\left[\Psi_o, \bar\Psi_o \right] = 
\left\langle\exp{\left(i\int d^d x \left[\bar {\mathcal{O}}^- \Psi_o^-
      + \bar\Psi_o^+ \mathcal{O}^+ \right] \right)}\right\rangle.
\end{equation}

Near the boundary of AdS,
$z = R^2/r \to 0$, the independent solutions of the 10-dimensional
Dirac equation are
\begin{eqnarray*}
\Psi(z,x) &\to& z^{\frac{d}{2} + m R} ~\Psi_+(x) 
            + z^{\frac{d}{2} - m R} ~\Psi_-(x) \\
\bar\Psi(z,x) &\to& z^{\frac{d}{2} + m R} ~\bar\Psi_+(x) 
            + z^{\frac{d}{2} - m R} ~\bar\Psi_-(x) .
\end{eqnarray*}
The solution
with $\Psi_-$ dominates near $z \to 0$,
thus $\Psi_- = \Psi_o$. The field $\Psi_-(x)$ acts
as a boundary source, and $\Psi_+(x)$ is the response function which
incorporates the quantum fluctuations.
The boundary sources for positive and negative chirality $\Psi_o^-$ and 
$\bar\Psi_o^+$ have dimensions $\frac{d}{2} - m R$. Consequently the
dimension of the CFT operators 
$\mathcal{O}^+$ and $\bar{\mathcal{O}}^-$ is $\frac{d}{2} + m R$.
Since the dimension of $\Psi_+$ is also $\frac{d}{2} + m R$, we expect that
$\Psi_+(x)$ is related to the expectation value of $\mathcal{O}$ in the
presence of the source $\Psi_o$:
$\Psi_+(x) \sim \left\langle {\mathcal{O}}\right\rangle_{\Psi_o}$.
Indeed $\left\langle {\mathcal{O}} \right\rangle
 =  (2 \Delta - d)~\Psi_+(x)$~\cite{Klebanov:1999tb}. 
Thus $\Psi_+$  acts as a
semiclassical field and is the boundary limit of the
normalizable string solution.

We consider first the classical solution dual to the valence Fock
state, described by the
massless 10-dimensional Dirac equation in
the bulk: $\Gamma^A D_A \hat \Psi  = 0$.
Full space  
coordinates are $x^A = ( x^\mu, z, y^a)$, with
$x^\mu$ the Minkowski variables and $z$ the
holographic coordinate. 
Coordinates in the compact space are $y^a$, and $g_{\perp a b}$ is the 
transverse metric. The full metric of spacetime is~\cite{Polchinski:2001tt}
\begin{equation}
ds^2 =
\frac{R^2}{z^2} e^{2A(z)} \left(\eta_{\mu \nu} dx^\mu dx^\nu - dz^2\right) 
       + g_{\perp a b} dy^a dy^b,
\label{eq:metric}
\end{equation}
where $A(z) \to 0$ as $z \to 0$, and behaves asymptotically as
a product of AdS space and a compact manifold $X$. 
A 10-dimensional field is represented by hat quantities: $\hat\Phi$, $\hat \Psi$; a
field on AdS space by  $\Phi$, $\Psi$, and 
$\phi$, $\psi$ represent fields in 4-dimensions.
Knowledge of the full geometry is required to solve the Dirac
equation. We truncate effectively the infrared
region at the end of AdS space at $z_o = 1 / \Lambda_{QCD}$, 
where string modes cannot propagate. We expand the
state $\hat \Psi$  in terms of spinors $\eta(y)$ of the Dirac
operator on a $d+1$ sphere with eigenvalues $\lambda_\kappa$ as
$\hat\Psi(x, z, y) = \sum_\kappa \Psi_\kappa(x, z) \eta_\kappa(y)$.
For each eigenvalue $\lambda$ the normalizable string modes are
\begin{equation}
\Psi(x,z) =  C e^{-i P \cdot x} z^{\frac{d + 1}{2}}  \left[
J_{\lambda R - \half}(z {\mathcal{M}} ) \mu_+(P) 
+ J_{\lambda R + \half} ( z {\mathcal{M}} ) \mu_-(P) \right] .
\label{eq:Diracmode}
\end{equation}
Four dimensional spinors
are  related by~\cite{Muck:1998iz} $\mu_- = (\Pslash/P) \mu_+$. 
The eigenvalues on $S^{d+1}$ are
$\lambda_\kappa R = \pm \left(\kappa + \frac{d}{2} + \half \right)$, 
$\kappa  = 0, 1, 2, ...$,
with multiplicity~\cite{Camporesi:1995fb}  
$D_{d+1}(\kappa) =  2^{\frac{d }{2}} \binom{d  + \kappa }{\kappa}~$.

The lowest dimension of a spin-$\half$ field on $AdS_{d+1}$ space is 
Dim$[\Psi] = d + \half$.
A spin-$\half$ field in the d-dimensional boundary theory has
dimensions  Dim$[\psi] = (d -1)/2$.
It is quite remarkable that a dual CFT boundary operator $\mathcal{O}$ 
with dimension $d + \half$ can
only be constructed for d = 2, 4 and d $\to \infty$
with  a product of five, three and two quark fields
respectively, to match the string dimensions in the bulk.

For $d = 4$, the spinor irreps are
$\bf 4$, $\bf 20$, $\bf 60$, $\bf 140$, \dots
Classical spin-$\half$ string solutions are labeled by eigenvalues 
of the Dirac operator on $X$. 
The lowest string mode for $\kappa = 0$ has dimension  $\Delta =
\frac{9}{2}$, and transforms as a $\bf 4$ under the $SU(4)$ $R$-symmetry.
The corresponding CFT operator $\mathcal{O}_\frac{9}{2}$ is constructed
as the product of three quark fields $\psi^r$
transforming as a $\bf\bar 4$ of $SU(4)$,   
since ${\bf \bar 4} \otimes \bf {\bar 4} \otimes {\bf \bar 4} 
\to {\bf 4}$. The new degeneracy could be interpreted as a flavor
symmetry with $r = u,d,c,s$, which is broken by quark masses.
With respect to color
$\mathcal{O}_{9/2}$ has the gauge invariant form
$\mathcal{O}(x)_{9/2} 
= \psi_{\bf N_C}(x) \psi_{\bf N_C}(x) 
\psi_{\bf N_C(N_C -1)/2}(x)$,
where the representation $\bf N_C(N_C -1)/2$ follows from the
antisymmetric component of the tensor product ${\bf N_C} \otimes {\bf N_C}$.
For  $ N_C = 3$, we recover the usual form of the interpolating
 operator which creates a physical baryon in QCD(3+1): ~$\mathcal{O}_{9/2} = 
\epsilon_{a b c} \psi_a \psi_b \psi_c$.

QCD is fundamentally different from SYM
theories where matter fields appear in adjoint multiplets of $SU(N_C)$.
The $\mathcal{N} = 4$ 
theory is dual to the low energy
supergravity approximation to type IIB string~\cite{Maldacena:1997re} 
compactified on $AdS_5 \times S^5$. The SYM fields correspond to
closed strings and comprise a gluon field, six scalars, four
Majorana gluinos, and their antiparticles.
The introduction of quarks in the fundamental 
representation\footnote{The
  introduction of a finite number of $N_f$ branes is dual to the
  introduction of flavor~\cite{Karch:2002sh} with quarks in the
  fundamental representation, and leads to a calculable 
  spectrum~\cite{Kruczenski:2003be}.} is dual to
the introduction of an open string sector, where the strings end on a
brane and join together at a point in the AdS
geometry~\cite{Gross:1998gk}. The SYM particles are expected to acquire
a mass of the order of SUSY breaking scale and decouple from the theory.

The AdS/CFT correspondence is interpreted in the present context as a 
classical duality between
the lowest three-quark valence state in the asymptotic $3 + 1$
boundary and the lowest string mode in
$AdS_5 \times S^5$. Higher Fock components are a manifestation of the
quantum fluctuations of QCD and are conformal states in the limit of
massless quarks and vanishing QCD $\beta$-function~\cite{Brodsky:2004qb}. Metric
fluctuations of the bulk geometry about the fixed AdS background should
correspond to quantum fluctuations of Fock states above the valence state in
the limit where QCD appears nearly conformal.
As shown by Gubser, Klebanov and Polyakov, orbital excitations in the boundary  
correspond to string degrees of freedom propagating in the bulk
from quantum fluctuations in the AdS sector~\cite{Gubser:2002tv}.

Consider a gauge invariant interpolating operator which
creates an arbitrary Fock-state in the boundary
\begin{equation} \label{eq:IBO}
\mathcal{O}^{L}_n = 
{\rm Tr} \left( \psi D_{\ell_1} \psi D_{\ell_2} \psi \dots \bar\psi
  \psi \dots F D_{\ell_m} F \dots \right) ,
\end{equation}
where the $m$ derivatives determine the total spacetime orbital angular momentum, $L =
\sum_{i=1}^m \ell_i$. The conformal dimension of (\ref{eq:IBO}) is
$\Delta = \Delta_n + L$,
where $\Delta_n$ is the sum of the dimensions of the quarks,
antiquarks and gluons of the Fock component. An effective five-dimensional mass $\mu$
in the AdS wave equation corresponding to quantum
excitations about the fixed AdS metric is asymptotically
determined by spectral comparison of the string modes and the boundary operators
of QCD, while maintaining conformal invariance. Matching the dimension of the Fock
components at the Minkowski boundary we find the
relation  $\mu R =\hbar c L$. Thus, an $\ell$ quantum orbital at
the Minkowski boundary corresponds to a five-dimensional mass $\mu
\sim \hbar c~ \ell/R$ in
the bulk side. 
The four-dimensional mass spectrum $\mathcal{M}_L$ is determined by imposing boundary
conditions on one of the solutions of the Dirac equation
$\Psi^\pm(x, z_o) = 0$.
The solution of the spin-$\threehalf$ Rarita-Schwinger equation in AdS
space is more involved, but considerable simplification occurs for polarization
along Minkowski coordinates, $\Psi_\mu$, where it becomes
similar to the spin-$\half$ solution~\cite{Volovich:1998tj}.

\section{Baryon Spectrum}

~~~~~The study of the hadron spectrum is crucial for  our
understanding of bound states of strongly interacting relativistic
confined particles. Different QCD-based models often disagree, 
even in the identification of the relevant degrees of 
freedom~\cite{Capstick:2000qj,Klempt:2004yz}.
Studies of orbitally excited baryons based on the $1/N_C$ expansion
have been useful for identifying the
relevant effective operators and determine their relative 
importance~\cite{Goity:2003ab}.
An outstanding puzzle is that the spin-orbit splitting, which 
experimentally is very small, appear in the $1/N_C$
expansion as a zeroth-order effect. Recently, 
the computation of orbital excitations on the
lattice has been
extended up to spin-$\fivehalf$ states~\cite{Richards:2004xx}. 

The spectrum of $N$ and $\Delta$ baryon states is listed in Table \ref{baryontable}
according to total angular momentum-parity assignment given by the 
PDG~\cite{{Eidelman:2004wy}}.  To determine intrinsic spin 
and orbital momentum quantum numbers we
have used the conventional $SU(6) \supset SU(3)_{flavor} \otimes
SU(2)_{spin}$ multiplet structure.
We limit ourselves to the light unflavored
hadron states and the introduction of massless quarks. 
Since  $m_{u,d} \ll \Lambda_{QCD}$, the light quarks
are extremely relativistic. Consequently
the mass of the hadrons corresponds essentially to the confined
kinetic energy of
massless quarks and gluons.
The intrinsic spin $S$ of a given hadron should match the spin of the dual
string.

\begin{table}
\begin{center}
{\begin{tabular}{@{}cccc@{}}
\hline
\multicolumn{4}{c}{}\\[-2ex]
 $SU(6)$ &   $S$ & $L$   &   Baryon State \\[0.5ex]
\hline
\multicolumn{4}{c}{}\\[-2ex]
${\bf 56}$ & $\half$ & 0  &  $N{\half^+}(939)$\\[0.5ex]
{}  & $\threehalf$& 0 &   $\Delta{\threehalf^+}(1232)$\\[1.5ex]
${\bf 70}$ & $\half$ & 1 &  $N{\half^-}(1535)~~ N{\threehalf^-}(1520)$ \\[0.5ex]
{}  & $\threehalf$ & 1 &  $N{\half^-}(1650)~~ N{\threehalf^-}(1700)~~N{\fivehalf^-}(1675)$\\[0.5ex]
{}  & $\half$ & 1  &  $\Delta{\half^-}(1620)~~
\Delta{\threehalf^-}(1700)$ \\[1.5ex]
${\bf 56}$ & $\half$ & 2  &  $N{\threehalf^+}(1720)~~ N{\fivehalf^+}(1680)$ \\[0.5ex]
{}  & $\threehalf$ & 2  &  $\Delta{\half^+}(1910)~~ \Delta{\threehalf^+}(1920)
                     ~~ \Delta{\fivehalf^+}(1905)~~\Delta{\sevenhalf^+}(1950)$\\[1.5ex]
${\bf 70}$ & $\half$ & 3 &  $N{\fivehalf^-}~~ ~~~~~~ N{\sevenhalf^-}$ \\[0.5ex]
{}  & $\threehalf$ & 3  &  $N{\threehalf^-}~~~~~~ ~~ N{\fivehalf^-}~~~~~~ ~~
                     N{\sevenhalf^-}(2190)~~ N{\ninehalf^-}(2250)$\\[0.5ex]
{}  & $\half$ & 3  & $\Delta{\fivehalf^-}~~~~~~ ~~ \Delta{\sevenhalf^-}$ \\[1.5ex]
${\bf 56}$ & $\half$ & 4 &  $N{\sevenhalf^+}~~~~~~ ~~ N{\ninehalf^+}(2220)$ \\[0.5ex]
{}  & $\threehalf $ & 4 &  $\Delta{\fivehalf^+}~~~~~~ ~~ \Delta{\sevenhalf^+} ~~~~~~ ~~
                       \Delta{\ninehalf^+}~~~~~~ ~~\Delta{\elevenhalf^+}(2420)$\\[1.5ex]
${\bf 70}$ & $\half$ & 5 &  $N{\ninehalf^-}~~~~~~ ~~ N{\elevenhalf^-}$ \\[0.5ex]
{}  & $\threehalf$ & 5 &  $N{\sevenhalf^-}~~~~~~~ ~~ N{\ninehalf^-}~~~~~~ ~~
      N{\elevenhalf^-}(2600)~~ N{\thirteenhalf^-} $\\[1.25ex] 
\hline\hline
\end{tabular}}
\caption{$SU(6)$ multiplet structure for the known $N$ and
$\Delta$ baryon resonances including internal spin $S$ and orbital 
angular momentum $L$
quantum numbers. Radial excitations are non included in the table.} 
\label{baryontable}
\end{center}
\end{table}

\begin{figure}
\centering
\includegraphics[angle=0,width=9cm]{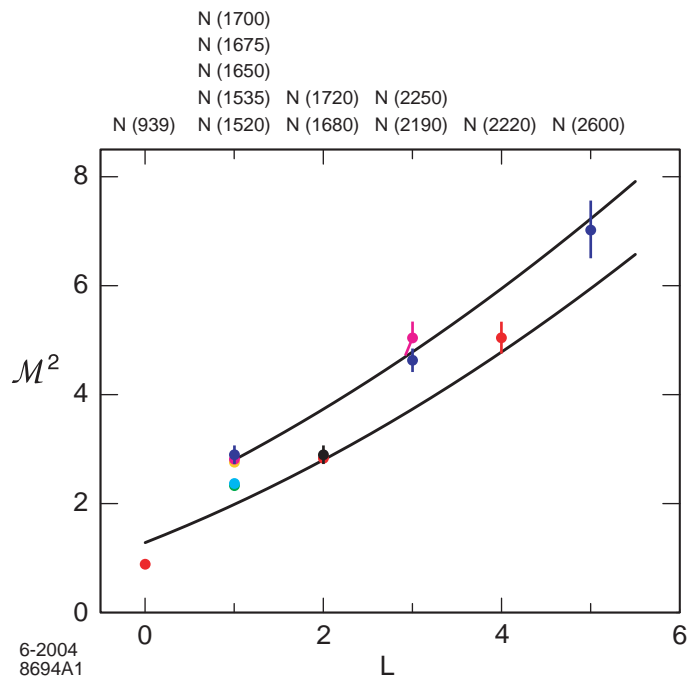}
\caption{Nucleon orbital spectrum for a value of $\Lambda_{QCD}$ = 0.22 GeV. 
The lower curve corresponds to nucleon
  states dual to spin-$\half$ string modes in the bulk. The upper curve corresponds to nucleon
states dual to string-$\threehalf$ modes.} 
\label{fig:Nucleonspec}
\end{figure}

\begin{figure}
\centering
\includegraphics[angle=0,width=9cm]{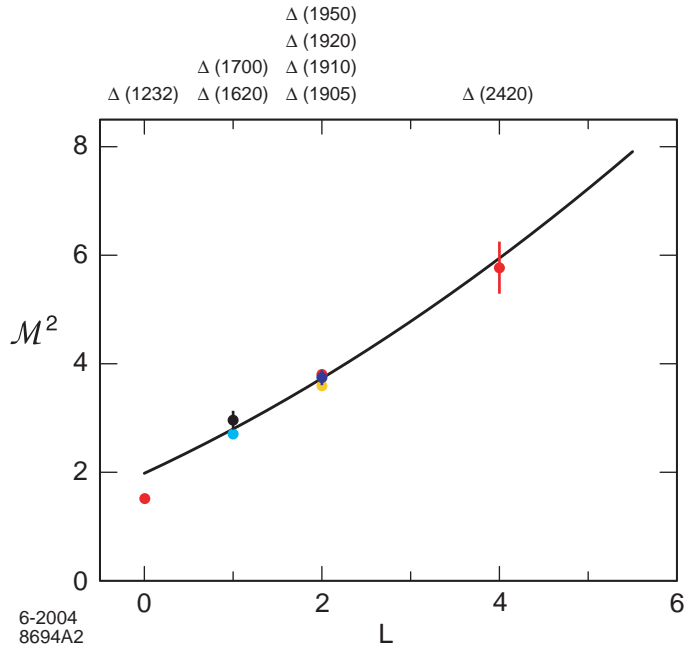}
\caption{Delta orbital spectrum for $\Lambda_{QCD}$ = 0.22 GeV.
 The Delta states dual to spin-$\half$ and
 spin-$\threehalf$ string modes in the bulk lie on the same
 trajectory.}
\label{fig:Deltaspec}
\end{figure}

We present in Fig. \ref{fig:Nucleonspec} the orbital spectrum of the
nucleon states and in Fig. \ref{fig:Deltaspec} the $\Delta$
orbital resonances. We plot the values of $\mathcal{M}^2$
as a function of $L$. The nucleon states with intrinsic spin $S = \half$ lie on a curve
below the nucleons with $S = \threehalf$.
We have chosen our boundary conditions by
imposing the condition $\Psi^+(x, z_o) = 0$ on the positive
chirality modes for $S = \half$ nucleons, and $\Psi^-_\mu(x, z_o) = 0$ on
the chirality minus strings for $S =\threehalf$. In contrast to the
nucleons, all of the know $\Delta$ orbital states with 
$S = \half$ and $S = \threehalf$ lie on the same trajectory. The boundary
conditions in this case are imposed on the chirality minus string modes.
The numerical solution corresponding to the roots of Bessel functions in
(\ref{eq:Diracmode}),
give the nonlinear trajectories indicated in the figures. All the
curves correspond to the value $\Lambda_{QCD} = 0.22$ GeV,
which is the only actual parameter aside from the choice of the
boundary conditions.
The results for each trajectory show a clustering of states with
the same orbital $L$,
consistent with strongly suppressed spin-orbit forces; this is a severe problem
for QCD models using one-gluon exchange. The  results also indicate a
parity degeneracy between states in the parallel trajectories shown in
Fig. \ref{fig:Nucleonspec}, as seen by displacing the
upper curve by one unit of $L$ to the right. Nucleon states
with $S = \threehalf$ and $\Delta$ resonances fall on the same 
trajectory~\cite{Klempt:2004yz}.

A string wavefunction with a node in the
holographic coordinate $z$ should correspond to a radial baryonic
resonance with a
node in the interquark separation, such as the positive parity Roper 
state $N\half^+(1440)$.
The first radial AdS eigenvalue corresponds to a mass around 1.85 GeV 
which would rather agree with the higher $N\half^+(1710)$ state. Recent
lattice results~\cite{Dong:2003zf} are consistent with the  
$N\half^+(1440)$ as a radial excited state.
However, a possible interpretation as a $N\half^+(1710)$ cannot be ruled 
out by present extrapolations. Lattice simulations at lower quark
mass should be undertaken to have a definite answer.

\section{Concluding Remarks}

~~~~~We have described aspects of the $\mathcal{N} = 4$ SYM orbital baryon spectrum
introducing an open string sector and a confining background by effectively cutting
AdS space in the far infrared AdS region at $r_o = \Lambda_{QCD} R^2$.
Since only one parameter, the QCD
scale $\Lambda_{QCD}$, is used, the agreement of the model with   
the pattern of the physical light baryon spectrum is remarkable. 
This agreement possibly reflects the fact that our analysis 
is based on a conformal template, which is a good initial 
approximation to QCD~\cite{Brodsky:2004qb}. We have chosen a special representation
to construct a three quark baryon, and the results are effectively
independent of $N_C$. This is consistent with results from lattice
gauge theory for glueballs~\cite{Teper:2001ja} where very little 
dependence on $N_C$  at small
lattice spacing is found for $N_C > 3$.
The  gauge/string
correspondence presented here appears as a powerful organizing
principle to classify and compute the mass eigenvalues of 
baryon resonances. A better
understanding of nonconformal aspects of the metric and
the nature of quantum fluctuations
about the AdS geometry is required. 
Our results suggest that fundamental features of the hadron
spectrum and QCD can be understood in terms of the nature of a higher
dimensional dual gravity theory.
Further discussion of some of theses
issues including a computation of the meson spectrum will be given elsewhere.

\section{Acknowledgements}

~~~~~We thank Oleg Andreev, Simon Capstick, Lance Dixon, Jose Goity, Robert Jaffe, Igor
Klebanov, David Mateos, David Richards, Jorge Russo and Matt Strassler 
for helpful comments.

\vspace{0.8cm}

\noindent{\bf \large Added Note}

\smallskip

After completion of this work we have learned of several studies
related to the  present discussion: (1) A  mapping of meson operators to
fluctuations of the supergravity background 
has been conjectured  in the framework  
of~\cite{Karch:2002sh} in the low energy limit of closed strings.
See: J. Erdmenger and I. Kirsh, arXiv:hep-th/0408113. (2) According to a recent
proposal the lowest
trajectory of Fig. \ref{fig:Nucleonspec} correspond to ``good''
diquarks, and the upper to ``bad'' diquarks . All the states in
Fig. \ref{fig:Deltaspec} correspond to ``bad'' diquarks. See:
F.~Wilczek, arXiv:hep-ph/0409168. (3) A recent lattice study of the
Roper Resonance points out the necessity to carry further
simulations at lower quark masses. See: D. Guadagnoli, M. Papinutto and S. Simula,
arXiv:hep-lat/0409011.

\end{document}